\documentclass[a4paper]{article}
\pdfoutput=1

\usepackage[utf8]{inputenc}
\usepackage{multirow}
\usepackage{amssymb}
\usepackage{makecell}
\usepackage{rotating}
\usepackage{bytefield}
\usepackage{url}
\usepackage{xcolor}
\usepackage{tabularx}
\usepackage{booktabs}
\usepackage{caption}
\usepackage{ragged2e}
\usepackage{authblk}
\usepackage{float}
\usepackage{listings}

\usepackage{enumitem}

\newcolumntype{Y}{>{\RaggedRight\arraybackslash}X}

\definecolor{brickred}{rgb}{0.8, 0.25, 0.33}

\newcommand{\bitlabel}[2]{\bitbox[]{#1}
{\raisebox{0pt}[0ex][0pt]{\turnbox{0}{\fontsize{8}{8}\selectfont#2}}}}

\newcommand{\tbltype}[1]{\multirow{1}{*}{\makecell[l]{#1}}}

\begin{document}
\title{Practical Cryptographic Data Integrity\\Protection with Full Disk Encryption\\Extended Version}

\author[1,2]{Milan Brož}
\author[1]{Mikuláš Patočka}
\author[2]{Vashek Matyáš}
\affil[1]{Red Hat Czech, \texttt{\{mbroz,mpatocka\}@redhat.com}}
\affil[2]{Masaryk University, Faculty of Informatics, Czech Republic\\
\texttt{\{xbroz,matyas\}@fi.muni.cz}}

\date{}

\maketitle

\begin{abstract}
Full Disk Encryption (FDE) has become a widely used security feature.
Although FDE can provide confidentiality, it generally does not provide
cryptographic data integrity protection.
We introduce an algorithm-agnostic solution that provides both
data integrity and confidentiality protection at the disk sector layer.
Our open-source solution is intended for drives without any special hardware
extensions and is based on per-sector metadata fields implemented in software.
Our implementation has been included in the Linux kernel since the version~4.12.
This is extended version of our article that appears in \emph{IFIP SEC 2018} conference
proceedings~\cite{fde_ifipsec18}.
\end{abstract}

\section{Introduction}\label{sec:intro}
Storage encryption has found its way into all types of data
processing systems -- from mobiles up to big data warehouses.
Full Disk Encryption (FDE) is a way to not only provide data
confidentiality but also provide an easy data disposal procedure
(by shredding the used cryptographic key).
FDE and the underlying block device work with the disk sector as an atomic and
independent encryption unit, which means that FDE can be transparently placed
inside the disk sector processing chain.

A major shortcoming of current FDE implementations is the \emph{absence of data
integrity protection}.
Confidentiality is guaranteed by symmetric encryption algorithms, but the nature
of length-preserving encryption (a plaintext sector has the same size as
the encrypted one) does not allow for any metadata that can store
integrity protection information.

Cryptographic data integrity protection is useful not only for detecting random
data corruption~\cite{krioukov2008parity}
(where a CRC-like solution may suffice) but also for providing a countermeasure
to targeted data modification attacks~\cite{turpe2009attacking}.
Currently deployed FDE systems provide no means for proving that data were
written by the actual user. An attacker can place arbitrary data on the storage
media to later harm the user.
FDE systems such dm-crypt~\cite{nmihde,dmcrypt} or
BitLocker~\cite{Elephant} simply ignore the data integrity problem, and
the only way to detect an integrity failure is the so-called
\emph{poor man's authentication} (user can recognize data from
garbage produced by the decryption of a corrupted ciphertext).

The aim of our work is to \emph{demonstrate that we can build practical
cryptographic data integrity and confidentiality protection} on the disk sector
layer with acceptable performance and without the need for any special hardware.
Our solution is an open-source extension of existing tools; we do not invent
any new cryptographic or storage concepts.
To achieve this goal, we implemented a per-sector metadata store over
commercial off-the-shelf (COTS) devices.
We provide an \emph{open-source implementation as a part of the mainline
Linux kernel}, where it is crucial to avoid proprietary and patented technology.

We focus on the \emph{security of authenticated encryption and the
algorithm-agnostic implementation}.
Our main contributions are as follows:
\begin{itemize}[noitemsep,topsep=0pt]
\item separation of storage and cryptographic parts that
allow changing the underlying per-sector metadata store implementation without
modifying the encryption layer,
\item the concept and implementation of emulated per-sector metadata,
\item algorithm-agnostic implementation of sector authenticated encryption
in the Linux kernel and
\item use of random initialization vector for FDE.
\end{itemize}

Storage security has often been perceived as an additional function that can
easily be added later. The history of length-preserving FDE is a demonstration
of this false idea. Although it is a simple application of cryptography concepts,
some vendors deployed FDE not only with known vulnerabilities but also
with incorrectly applied cryptography algorithms
~\cite{eprint-2015-26408,Saarinen2005}.

The remainder of this paper is organized as follows.
Section~\ref{sec:usermodels} discusses the threat model and introduces
the data corruption problem.
Section~\ref{sec:methods} describes proposed algorithms for encryption
and integrity protection.
Section~\ref{sec:metadata} discusses how to store additional integrity data
and extends the proposed metadata store concept to construct a reliable virtual
device that handles both data and metadata.
Sections~\ref{sec:implementation} and~\ref{sec:performance} describe
our practical implementation and performance evaluation.
Section~\ref{sec:conclusion} concludes our paper and identifies future work.

\section{Threat Model and Use Cases}\label{sec:usermodels}
Use cases for of FDE can be categorized to several situations, like
\begin{itemize}[noitemsep,topsep=0pt]
\item \emph{stolen hardware} (mobile device, laptop),
\item \emph{devices in repair},
\item resold \emph{improperly wiped} devices,
\item \emph{virtual device} in a multi-tenant environment, or
\item a \emph{mobile device} storage.
\end{itemize}

In all of these scenarios, confidential data can leak out of the control of
the owner. A recent analysis of the content of used drives~\cite{uae-disks2016}
shows that encryption is often still not used despite the importance
of storage data encryption being recognized for a long time~\cite{riedel2002framework}.

Our threat model adds to all these scenarios the detection of \emph{data tampering}
on leaked (not-wiped) devices  and expects that an attacker has limited ability
to record device changes with access snapshots of the device in time.
The model recognizes the \emph{Focused Opportunistic} attacker defined in Table~\ref{tbl:attackers}.
It does not protect data in situations where a device is in active (unlocked) state
or an attacker can get the unlocking passphrase (and thus encryption key) directly.
As mentioned in Section~\ref{sec:intro}, our model expects COTS devices, it cannot
rely on use of any of tamper-proof cryptographic devices like Hardware Security Modules (HSMs).

\subsection{Attackers}
We define three simplified types of attackers, as summarized in Table~\ref{tbl:attackers}.
The most common type of attacker for FDE is a \emph{random} attacker.
We define a \emph{focused opportunistic} attacker type for sophisticated
attacks that focus on the situation where the stolen device is returned to the user.
In the \emph{targeted} attacker case, FDE will not provide sufficient
protection without additional countermeasures.

\begin{table}[h]
\def\arraystretch{1.2}
\vspace{6pt}
\begin{tabularx}{\linewidth}{@{} lY @{}}
\toprule
\textbf{Type} & \textbf{Description} \\
\midrule
\tbltype{Random}  &
  An attacker with user skills, can investigate disk content.
  Focuses on hardware, to steal and sell it. Valuable data are simply a bonus.
  A~typical representative for this category is a random thief. \\
\tbltype{Focused\\Opportunistic} &
  An attacker can run advanced forensic tools and can
  use documented attacks (such as a brute-force password search).
  The goal is to obtain data, analyze them, and use them to make a profit.
  The attack is not targeted to a specific user, but if the recovered data
  allow such an attack, the attacker can switch to this type of attack.
  A typical representative is a computer technician with access to devices
  for repair.
  In some cases, the attacker can access the device repeatedly,
  but only in limited opportunistic occasions. \\
\tbltype{Targeted} &
  A top skilled attacker that uses the best possible attack vectors, typically
  to focus on a specific user with exactly defined goals in advance.
  The user system is generally under full attacker control. \\
\bottomrule
\end{tabularx}
\caption{Attackers.}
\label{tbl:attackers}
\end{table}

\begin{table}[h]
\def\arraystretch{1.2}
\centering
\begin{tabularx}{\linewidth}{@{} lY @{}}
\toprule
\textbf{Type} & \textbf{Description} \\
\midrule
\tbltype{Pure FDE} &
  Length-preserving encryption that provides confidentiality only. \\
\tbltype{Authenticated\\FDE} &
  Encryption that provides both confidentiality and integrity protection, but
  limited by COTS devices (no hardware for authentication). \\
\tbltype{HW-trusted} &
  The ideal solution with confidentiality and integrity protection.
  It~stores some additional information to external trusted storage in such a way
  that the system can detect data replay. \\
\bottomrule
\end{tabularx}
\caption{Discussed types of FDE protection.}
\label{tbl:fdetypes}
\end{table}

\newcommand{\yes}{\checkmark}
\newcommand{\no}{$\times$}
\newcommand{\maybe}{(\yes)}
\begin{table}[h]
\centering
\vspace{6pt}
\def\arraystretch{0.9}
\begin{tabular}{ccccc}
\toprule
\textbf{\makecell{FDE type:}}&
\textbf{\makecell{None}} &
\textbf{\makecell{Pure\\FDE}} &
\textbf{\makecell{Auth.\\FDE}} &
\textbf{\makecell{HW\\trusted}} \\
\midrule
\makecell{Confidentiality}             & \no  & \yes   & \yes   & \yes   \\
\makecell{Integrity}                   & \no  & \no    & \yes   & \yes   \\
\makecell{COTS hardware}               & \yes & \yes   & \yes   & \no    \\
\makecell{Detect silent corruption}    & \no  & \no    & \yes   & \yes   \\
\makecell{Detect data tampering}       & \no  & \no    & \yes   & \yes   \\
\makecell{Detect data replay}          & \no  & \no    & \no    & \yes   \\
\makecell{Whole sector change}         & \no  & \maybe & \maybe & \maybe \\
\bottomrule
\end{tabular}
\caption{Overview of FDE features.}
\label{tbl:protections}
\end{table}

\subsection{FDE Protection Types}
For the description of our model, we define three basic levels of protection, as
summarized in Table~\ref{tbl:fdetypes}.
Here, FDE protection (of any type) means that the data confidentiality
is enforced.
A simple case of a device theft means that only hardware is lost.
Data remain encrypted and not accessible to the attacker. This scenario
is covered by the \emph{Pure FDE} protection.
The importance of \emph{authenticated FDE} comes into play when the stolen
or seized device returns to the user (and this often occurs in reality; and example
can be mandatory border checks).

This situation is generally enforced by a security policy and
compliance of users. The reality is different -- experiments
show that people plug-in foreign devices, even if such devices are obtained under
suspicious circumstances~\cite{7546509}.

Authenticated encryption enforces that a user cannot read tampered data
but will see an \emph{authentication error}.
It not only stops any attempts to use tampered data on higher layers,
but also helps a user to realize that the device is no longer trustworthy.
An overview of the features among FDE types is summarized in
Table~\ref{tbl:protections}.

\subsection{Data Corruption and Forgery}\label{sec:dataenc}
So-called silent data corruption~\cite{Bairavasundaram:2008:ADC:1416944.1416947}
is a common problem in persistent storage.
This problem occurs when data are unintentionally and randomly corrupted while
traversing through the storage stack. It is generally caused by flaky hardware
(unstable memory bits and loose cables), triggered by an external influence
(ionizing radiation) or by misplacement of data (correct data are written to
an incorrect place).
Data are then stored and used in a corrupted form.

A solution is to detect data corruption by checksums such as CRC32~\cite{rfc3385}.
This solution is not adequate if we want to detect an intentional
unauthorized change (an attacker will simply fix a checksum).

An active attacker can not only cause undetected data corruption by simulating
silent data corruption but can also attempt to forge the data by corrupting
a specific disk area.
A more sophisticated attacker can store additional data using
steganographic techniques (conceal other data) to unused areas of a disk.
We have to use a cryptographic integrity protection to detect such a situation.

\subsection{Replay Attacks}\label{sec:replay}
In the strong view of cryptographic integrity protection, we should also detect
data replacement using old content (revert to snapshots, also called
\emph{replay attack}~\cite{vanDijk:2007:OUS:1314354.1314364}).
Such a requirement cannot be fulfilled without an additional trusted metadata
store independent of the storage itself. If we have such a storage, it can be used
to store a Tamper Evident Counter (TEC).

The content of the entire storage can always be completely replaced with an older snapshot.
Without additional and trusted information, users cannot recognize such a situation.
An attacker can also revert only a partial part of the storage (in our case,
selected sectors).
From the cryptographic perspective, this situation cannot be completely
prevented or detected (it would require breaking the independence of sectors).

In this text, we present algorithms that \emph{do not protect} from the replay
attack.
This decision is based on the fact that our work is focused on utilizing
standard disk drives without any additional hardware requirements.

\section{Encryption and Data Integrity}\label{sec:methods}
Data encryption and integrity protection can be performed on different layers.
An efficiency advantage comes with implementation on higher layers (only
used areas are processed). However, integrity protection on the FDE layer
provides at least some integrity protection of storage space, even in situations
when a higher layer does not provide such guarantees (unfortunately, this is
still quite common).

\subsection{Length-preserving Encryption}
Length-preserving encryption algorithms are used in current FDE solutions.
These algorithms transform original data (plaintext) to its encrypted form
(ciphertext), and confidentiality is based on the use of secret keys.
Storage data are (for performance reasons) always encrypted with symmetric
block ciphers.

The block sizes of these symmetric ciphers are typically 16 bytes.
The device sector size is considerably larger (at least 512~bytes);
thus, to apply encryption per sector, we have to utilize
a \emph{block encryption mode} inside a sector.
Current FDE systems use CBC~\cite{Elephant} or
XTS~\cite{Dworkin:2010:SRB:2206252} modes.
The CBC mode has many known problems~\cite{nmihde} and specific requirements
(see the note in Section~\ref{sec:iv}).

\label{sec:xts}The XTS mode, due to the internal parallel processing
of blocks, can leak more information about a change in the plaintext than other
modes. If only a single byte changes in a sector, then we can localize the change
with the internal cipher block granularity. An ideal FDE system should produce
a pseudo-random change of the entire sector data.
Also, these modes produces the same plaintext data encrypted
to the same sector always produce the same ciphertext.
In cryptography, this means that such a system does not fully provide
indistinguishability under chosen plaintext attack (IND-CPA)~\cite{Gjosteen2005,Khati2017}.

\subsection{Authenticated Encryption}
We have two options for integrity protection combined with device
encryption: either to use \emph{Authenticated Encryption with Additional
Data} (AEAD)~\cite{Bellare:2008:AER:1410264.1410269,Rogaway:2002:AA:586110.586125}
or to combine length-preserving encryption with an additional cryptographic
integrity operation. The major difference is that for the combined mode, we can
ignore integrity tags and decrypt the data without such tags. In the AEAD mode,
the authentication is an integral part of decryption. Additionally,
for the combined mode, we need to provide two separate keys (encryption
and authentication), whereas the AEAD mode generally derives the authentication
key internally.
Both mentioned integrity methods calculate an \emph{authentication tag} from
the final ciphertext (encrypt-then-MAC).

\begin{table}[h]
\centering
\def\arraystretch{1.4}
\centering
\begin{tabularx}{\linewidth}{@{} lY @{}}
\toprule
\textbf{Mode} & \textbf{Description} \\
\midrule
\tbltype{AES-CBC} & AES~\cite{FIPS:2001:AES} non-authenticated mode used
  in legacy FDE systems.~\cite{nmihde} \\
\tbltype{AES-XTS} & AES~\cite{FIPS:2001:AES} non-authenticated mode used
  in recent FDE systems.~\cite{Dworkin:2010:SRB:2206252} \\
\tbltype{AES-GCM} & AES~\cite{FIPS:2001:AES} in the Galois/Counter
  authenticated mode.~\cite{Dworkin:2007:SRB:2206251}
  Due to only 96-bit nonce, it can be problematic in the FDE
  context.~\cite{bock2016nonce}\\
\tbltype{ChaCha20-Poly1305} &
  Authenticated mode based on the~ChaCha20~\cite{Bernstein_chacha,rfc7539} cipher and the~Poly1305 authenticator.\\
\tbltype{AEGIS128 / 256} & AEAD ciphers based on AES round function (CAESAR~\cite{caesar} finalist)~\cite{aegis}. \\
\tbltype{MORUS640 / 1280} & AEAD ciphers designed for modern hardware optimizations (CAESAR~\cite{caesar} finalist)~\cite{morus}. \\
  \bottomrule
\end{tabularx}
\caption{Examples of encryption algorithms.}
\label{tbl:encmodes}
\end{table}

The encryption operation output consists of the encrypted data
and the authentication tag.
Authentication mode with additional data (AEAD) calculates the authentication
tag not only from the input data but also from additional metadata, called
additional authentication data (AAD).
Table~\ref{tbl:encmodes} summarizes examples of the encryption modes mentioned
in this text.

\subsection{Initialization Vectors}\label{sec:iv}
The Initialization Vector (IV) is a value for an encryption mode that impacts
encryption. In FDE, the IV must always be derived from a sector number (offset
from the device start) to prevent malicious sector relocation.
The sector number guarantees that the same data written to different sectors
produce different ciphertexts.
The proper use of IVs and nonces depends on the exact encryption mode and is
\emph{critical} for the security of the entire solution.
Table~\ref{tbl:iv} briefly describes the IV types
used in our work.
For the CBC mode, we must use an IV that an adversary cannot predict;
otherwise, the IV value can be used (in combination with a specially formatted
plaintext) to create special patterns in the ciphertext
(watermarks)~\cite{nmihde,Saarinen2005}.
Some encryption modes (such as XTS) solve this problem by encrypting the IV
such that they can use a predictable sector number directly.
\clearpage
The IV must always be unique per sector. In some cases, the IV must be
a \emph{nonce} (a public value that is never reused).
Repeating an IV for different sectors not only opens a possibility to malicious
sector relocation but can also violate a security restriction (in the GCM mode
repeating a nonce value is fatal~\cite{cryptoeprint:2015:102}).

\begin{table}[h]
\def\arraystretch{1.2}
\centering
\begin{tabularx}{\linewidth}{@{} lY @{}}
\toprule
\textbf{IV} & \textbf{Description} \\
\midrule
plain64 & Sector as a 64-bit number (device offset).
          Used for the XTS mode~\cite{dmcrypt}. \\
ESSIV   & Encrypted Salt-Sector IV.
          The sector number is encrypted
          using a salt as the key.
          The salt is derived from the device key
          with a hash function. Used for the CBC mode~\cite{dmcrypt}. \\
random  & IV generated on every sector write from
          a Random Number Generator (RNG). Used for AEAD modes (see section~\ref{sec:ivrand}).\\
\bottomrule
\end{tabularx}
\caption{Initialization vectors (IV generators).}
\label{tbl:iv}
\end{table}

\subsection{Error Propagation in Encrypted Sector}
\begin{figure}[h]
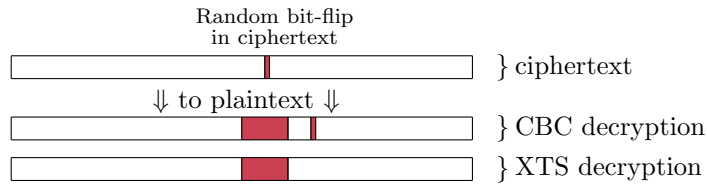

\centering
\vspace{5pt}
\begin{bytefield}[bitwidth=0.005\linewidth,bitheight=2ex]{100}
\bitlabel{80}{\makecell{Random bit-flip\\in ciphertext}} \\
\begin{rightwordgroup}{ciphertext}
  \bitbox{55}{} &
  \bitbox{1}{\color{brickred}\rule{\width}{\height}} &
  \bitbox{44}{}
\end{rightwordgroup} \\
\makecell{$\Downarrow$ to plaintext $\Downarrow$} \\
\begin{rightwordgroup}{CBC decryption}
  \bitbox{50}{} &
  \bitbox{10}{\color{brickred}\rule{\width}{\height}} &
  \bitbox{5}{} &
  \bitbox{1}{\color{brickred}\rule{\width}{\height}} &
  \bitbox{34}{}
\end{rightwordgroup} \\ [1ex]
\begin{rightwordgroup}{XTS decryption}
  \bitbox{50}{} &
  \bitbox{10}{\color{brickred}\rule{\width}{\height}} &
  \bitbox{40}{}
\end{rightwordgroup}
\end{bytefield}
\caption{Error propagation in encrypted sector.}
\label{fig:bit-error}
\end{figure}

With the symmetric encryption in the processing stack, error propagation is
amplified.
One bit flip causes a random corruption in at least one cipher block
(typically 16 bytes) up to the entire sector.
This is illustrated in Figure~\ref{fig:bit-error}.

Such a ``random corruption'' means that decrypted data are a product
of the decryption of a modified ciphertext.
By definition, modern ciphers~\cite{Ferguson:2010:CED:1841202} must produce
a random-looking output.
In other words, a user will see a block full of data garbage after decrypting
corrupted data.

\begin{figure}[h]
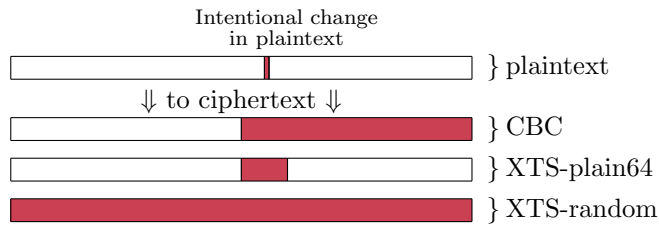

\centering
\vspace{8pt}
\begin{bytefield}[bitwidth=0.005\linewidth,bitheight=2ex]{100}
\bitlabel{80}{\makecell{Intentional change\\in plaintext}} \\
\begin{rightwordgroup}{plaintext}
  \bitbox{55}{} &
  \bitbox{1}{\color{brickred}\rule{\width}{\height}} &
  \bitbox{44}{}
\end{rightwordgroup} \\
\makecell{$\Downarrow$ to ciphertext $\Downarrow$} \\
\begin{rightwordgroup}{CBC}
  \bitbox{50}{} &
  \bitbox{50}{\color{brickred}\rule{\width}{\height}}
\end{rightwordgroup} \\ [1ex]
\begin{rightwordgroup}{XTS-plain64}
  \bitbox{50}{} &
  \bitbox{10}{\color{brickred}\rule{\width}{\height}} &
  \bitbox{40}{}
\end{rightwordgroup} \\ [1ex]
\begin{rightwordgroup}{XTS-random}
  \bitbox{100}{\color{brickred}\rule{\width}{\height}}
\end{rightwordgroup} \\ [1ex]
\end{bytefield}
\caption{Change propagation to encrypted sector.}
\label{fig:bit-change}
\end{figure}

For encryption, the change propagation is, in fact, a desirable effect.
The ideal situation is that any change in the sector data is propagated to
a pseudo-random change to the whole encrypted sector as illustrated
in Figure~\ref{fig:bit-change}.

\section{Metadata Storage Placement}\label{sec:metadata}
The absence of per-sector metadata (to store integrity protection data) is
a well-known problem.
Integrity protection requires a length expansion of the processed data and
thus needs additional storage~\cite{Sivathanu:2005:EDI:1103780.1103784}.
Common sector sizes are 512 and 4096~bytes, and we need an independent metadata
per-sector.

\subsection{Metadata in Hardware Sector}
A reliable way to handle integrity metadata is to do so directly
in the device hardware sector.
An in-sector integrity data approach appeared in 2003
as the T10 Data Integrity Field (DIF) extension for SCSI
devices~\cite{dif-justification}.
This idea was implemented by vendors as T10 Protection Information (T10-PI) where
a sector is expanded by 8~bytes (the sector size is 520 bytes)~\cite{petersen}.
For COTS devices, the DIF extension is quite rare, expensive and requires
a special controller.
The fixed provided metadata space is not large enough for cryptographic data
integrity protection that requires storing an Initialization Vector to metadata.

\subsection{Metadata Stored Separately}
Per-sector metadata can be stored in a separate storage space on the same
disk or an external fast storage~\cite{Sivathanu:2005:EDI:1103780.1103784}.
We can also join several smaller hardware sectors and metadata into one large
virtual sector presented to the upper layer.
This approach is used in authenticated modes in the FreeBSD GEOM encryption
system~\cite{Kamp:2003:GGB:1250972.1250979,geli}.
The initial GEOM Based Disk Encryption (GBDE)~\cite{Kamp:2003:GGB:1250972.1250979}
tried to use additional metadata for generated per-sector keys. This design
does not provide safe atomic sector updates and also significantly decreases
performance (throughput is only 20-25\% of an underlying device)~\cite{gbde}.
Some of these problems were fixed by the GELI disk encryption~\cite{geli}.
Here, the upper layer uses a 4096-byte sector, while
internally, it splits data into 512-byte native sectors (each contains its own
data and metadata).
In this case, for every presented sector, 8+1 native sectors are used.
This concept ensures that integrity data are handled on the atomic hardware
sectors.
Although the above filesystem sees the 4096-byte sector as an atomic unit,
the device operates with 512-byte units and can possibly interleave multiple
writes.
Another problem with this design is that the virtual sector is
always larger than the native device sector. If the native sector is 4096~bytes,
then the presented virtual sector can be larger than the optimal
block size for the filesystem above (for some filesystems, the optimal size can
be a page size, and thus, it is 4096~bytes in most situations).

Enterprise storage vendors also implement a multiple-sector schema,
generally with 8~data sectors + 1~common metadata
sector~\cite{Bairavasundaram:2008:ADC:1416944.1416947}.

\subsection{Interleaved Metadata Sectors}
\begin{table}[h]
\centering
\def\arraystretch{1.1}
  \begin{tabular}{crr}
  \toprule
   \makecell{\textbf{\makecell{Metadata}}\\\lbrack bytes\rbrack} &
   \makecell{\textbf{Space}~\lbrack\%\rbrack\\512B sector} &
   \makecell{\textbf{Space}~\lbrack\%\rbrack\\4096B sector} \\
   \midrule
    4 (32 bits)  &  0.78 & 0.10 \\
   16 (128 bits) &  3.03 & 0.39 \\
   28 (224 bits) &  5.26 & 0.68 \\
   32 (256 bits) &  5.88 & 0.78 \\
   48 (284 bits) &  9.09 & 1.16 \\
   64 (512 bits) & 11.11 & 1.54 \\
   80 (640 bits) & 14.29 & 1.92 \\
   \bottomrule
  \end{tabular}
\caption{Space for per-sector metadata.}
\label{tbl:metaspace}
\end{table}

Our solution combines the use of a device integrity profile with a metadata
per-sector stored independently in dedicated sectors. The presented sector
size is configurable. We can use the same size as a native device, but we can
also increase the presented sector size (atomicity is then ensured
by journaling, as described in Section~\ref{sec:journal}).
The combined data and metadata are then presented to the block layer as a new
virtual device with a specific integrity profile (software-emulated
DIF-enabled device).

\begin{figure}[h]
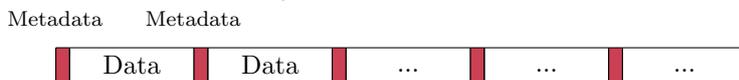

\centering
\begin{bytefield}[bitwidth=0.015\linewidth,bitheight=3ex]{50}
\bitlabel{-7}{\makecell{Metadata}}
\bitlabel{27}{\makecell{Metadata}} \\
\bitbox{1}{\color{brickred}\rule{\width}{\height}} & \bitbox{9}{Data} &
\bitbox{1}{\color{brickred}\rule{\width}{\height}} & \bitbox{9}{Data} &
\bitbox{1}{\color{brickred}\rule{\width}{\height}} & \bitbox{9}{\makecell{...}} &
\bitbox{1}{\color{brickred}\rule{\width}{\height}} & \bitbox{9}{\makecell{...}} &
\bitbox{1}{\color{brickred}\rule{\width}{\height}} & \bitbox{9}{\makecell{...}}
\end{bytefield}
\caption{Interleaved metadata sectors.}
\label{fig:interleaved-sectors}
\end{figure}

The metadata are stored in special sectors that are interleaved with data
sectors. According to the required metadata size, one sector contains
a metadata store for several consecutive sectors.
An illustration of this layout is provided in Figure~\ref{fig:interleaved-sectors}.
Required space examples are illustrated in Table~\ref{tbl:metaspace}.
The intereleaving format easily allows to resize block device later (in 
specific steps according to interleaving parameters).

\noindent
The required size of additional sectors is calculated as follows:
\begin{center}
\large $TagsPerSector = \lfloor\frac{SectorSize}{TagSize}\rfloor$

\large $Tags_{sectors} = \lceil\frac{Data_{sectors}}{TagsPerSector}\rceil$
\end{center}

The use of storage is optimal (no wasted space) when the sector size is
a multiple of the tag size.
Additional metadata space (part of a device used for metadata) is calculated as
\begin{center}
\Large $100~\frac{Tags_{sectors}}{Data_{sectors} + Tags_{sectors}}~~\%$.
\end{center}

\subsection{Recovery on Write Failure}
A device must provide atomic updating of both data and metadata.
A situation in which one part is written to media while another part failed
must not occur.
Furthermore, metadata sectors are packed with tags for multiple sectors;
thus, a write failure must not cause an integrity validation failure for other
sectors.

A metadata-enabled device must implement a data journal that provides
a reliable recovery in the case of
a power failure.\label{sec:journal} The journal can be switched off
if an upper layer provides its own protection of data.

In some specific situations, a journal could provide unintended additional
information about the encrypted device (such as last written data offsets or
old data content of unfinished sector writes).
A journal requires additional storage and decreases performance on write
(data are written twice). The journal size is generally just a fraction
of the device capacity.

\section{Linux Kernel Implementation}\label{sec:implementation}
We split the solution into \emph{storage} and \emph{cryptographic} parts.
The former provides an area for per-sector metadata for commercial
off-the-shelf (COTS) storage devices. The latter extends disk encryption
by cryptographically sound integrity protection.
For key management, we extended the LUKS encryption system~\cite{luks} to provide
support for our integrity protection.

Our approach is based on the existing block layer integrity profile
infrastructure~\cite{petersen} and is fully contained in the device-mapper kernel
subsystem.
The only condition is that the Linux kernel must be compiled with
support for the block integrity profile infrastructure
(\texttt{CONFIG\_BLK\_DEV\_INTEGRITY} option).

The implementation is fully contained in the device-mapper kernel
subsystem and consists of a new \emph{dm-integrity} target that provides
a virtual device with a configurable integrity profile and an extension
of \emph{dm-crypt} for using authenticated encryption.
Our implementation has been available in the mainline Linux kernel since
version 4.12~\cite{mainline_kernel}.

The principle of the device-mapper architecture is that separate functions are
combined by stacking several devices, each with a specific \emph{target} that
implements the needed function.

\noindent
For active integrity-enabled encryption, there are three
stacked devices:
\begin{itemize}[noitemsep,topsep=0pt]
\item \emph{dm-crypt device} (encrypts and authenticates data),
\item \emph{dm-integrity device} (provides per-sector metadata) and 
\item \emph{an underlying block device} (disk or partition).
\end{itemize}

The top-level device is available to a user and can be used
directly by an application or an arbitrary filesystem.
The \emph{dm-integrity} device can also operate in a standalone mode
(as described in Section~\ref{sec:dmintegrity}); in this case,
the \emph{dm-crypt} device is omitted.

\subsection{The dm-integrity Module}\label{sec:dmintegrity}
The implemented \emph{dm-integrity} device-mapper target creates a virtual
device with arbitrary-sized per-sector metadata over the standard block device
and presents it to the system as a device with a specific integrity profile.
The driver implements an optional data journal.

The \emph{dm-integrity} target can operate in two modes:
\begin{itemize}[noitemsep,topsep=0pt]
\item as a provider of per-sector metadata for the upper device
      (as a \emph{dm-crypt} target) or
\item in standalone mode, where it calculates and maintains basic data
      integrity checksums itself.
\end{itemize}

At the device initialization step, the integrity target takes the underlying
device and tag size and then writes the superblock to the device.
The following activations then just read data from the superblock.
An activated device creates a new integrity profile named \texttt{DM-DIF-EXT-TAG}.

The additional metadata are produced by the driver on top of the integrity
module (integrity module only processes metadata, it does not allocate them).
For reads, it combines the data and metadata sectors and submits them to
the upper layer.
For writes, it splits data and metadata and submits new I/O requests to
the underlying device.

\subsection{The dm-crypt Extension}
The \emph{dm-crypt} module is a Linux in-kernel implementation of FDE.
To support authenticated encryption, we implemented the following parts:
\begin{itemize}[noitemsep,topsep=0pt]
\item new parameter and configuration table format for integrity protection
      AEAD specification,
\item interface to the underlying I/O integrity profile\\(provided by
      \emph{dm-integrity}),
\item processing of requests specified in Section~\ref{sec:aead-sector},
\item new random-IV generator and
\item configurable encryption sector size.
\end{itemize}

All cryptographic primitives are provided by the kernel cryptographic API.
This provides us with all the hardware support and acceleration available on
a particular hardware platform.
The composed AEAD mode requires an independent key for integrity; thus, the
activation requires that the userspace provides a key that is internally split
into encryption and integrity parts.

When an integrity check fails, the entire I/O request is marked as failed with
the specific error code (EILSEQ) and returned to the block layer. To achieve
availability of data, a user can then stack redundant storage mapping (usually RAID)
above such an integrity protected device; this layer then should react to a detected
error by redirecting the operation to another redundant device.

The reality in current Linux RAID implementation is that it quietly amplifies
silent data corruption, as experiment with stacking dm-integrity device below
RAID shows~\cite{raid_fail}.
Here the adidtional authenticated layer actually stopped RAID mechanism from working
with corrupted data.

\subsection{Sector Authenticated Encryption}\label{sec:aead-sector}
To perform an authenticated encryption operation over a sector,
we define the format of a sector authentication request (Table~\ref{tbl:aead-request}).
The additional data (AAD) for our request contain the sector number and the IV.

Such a request detects a sector misplacement and a corrupted IV.
The request definition is compatible with the IEEE 1619 storage
standard~\cite{Dworkin:2007:SRB:2206251,4523925}.

\begin{table}[h]
\centering
\begin{tabular}{cccc}
\multicolumn{2}{c}{\textbf{AAD}}         & \textbf{DATA} & \textbf{AUTH} \\
\multicolumn{2}{c}{\small authenticated} & \small authenticated + encrypted &
                                           \textbf{TAG}  \\ \hline
\multicolumn{1}{|c|}{sector}             & \multicolumn{1}{c|}{IV} &
                                           \multicolumn{1}{c|}{data in/out} &
                                           \multicolumn{1}{c|}{tag} \\ \hline
\end{tabular}
\caption{AEAD sector authentication request.}
\label{tbl:aead-request}
\end{table}

\subsection{Random IV}\label{sec:ivrand}
We store persistent IV data with the additional per-sector metadata.
With the available extra metadata space, we define a new \emph{random-IV}.
The random-IV regenerates its value on every write operation by reading it from
a system Random Number Generator (RNG). The random-IV length should be at
least 128 bits to avoid collision (expecting that an attacker has the capability
to record all previous Initialization Vectors and sectors).

Since stored values are visible to an adversary, the random-IV cannot be used
for modes that require unpredictable IVs.

Some systems could be slow if the RNG is not able to produce this amount
of random data.
However, our tests in Section~\ref{sec:performance} with a recent kernel show
that the performance of the system RNG does not cause any noticeable problems.
The only limitation of the random-IV is that during the early boot
(before the RNG is initialized), the system must avoid writes to the device.
Systems typically boot from a read-only device, and the RNG
is properly seeded during this phase and becomes fully operational;
thus, it is not an issue.

For the XTS mode and additional authentication tag with a random IV, we
can improve change propagation as the entire sector is always pseudo-randomly
changed on each write, independently of the plaintext data content
(as illustrated in Figure~\ref{fig:bit-change}).
The random IV must not be used with length-preserving XTS mode (without
authentication tag) because in this mode decryption no longer depends on
sector position (sector relocation attacks are possible).

\subsection{Formatting the Device}
The integrity tag is calculated for all sectors, including sectors that are
not in use. On the initial activation, the device has to recalculate all
integrity tags; otherwise, all reads fail with an integrity error.
Such an integrity error can occur even on the initial device scan
(newly appeared device is automatically scanned for known signatures).

The initial device formatting is quite time consuming, but fortunately, it must
be done only once during the disk initialization phase. Disk encryption tools
should perform this step automatically.
We implemented such a function in the LUKS~\cite{luks} cryptsetup tool.

\begin{table}[h]
\centering
\def\arraystretch{1.4}
  \begin{tabular}{ccc}
  \toprule
   \multirow{2}{*}{\textbf{\makecell{Encryption\\and\\Integrity Mode}}} &
   \multicolumn{2}{c}{\textbf{Metadata}} \\ &
                   \makecell{IV\\\lbrack bytes\rbrack} &
                   \makecell{TAG\\\lbrack bytes\rbrack} \\
   \midrule
   \makecell{NULL cipher: no encryption, no integrity} & -  & -  \\
   \makecell{no encryption, CRC32 integrity}           & -  & 4  \\
   \makecell{AES256-XTS-plain64, no integrity}         & -  & -  \\
   \makecell{AES256-XTS-random, no integrity}          & 16 & -  \\
   \makecell{AES256-GCM-random, AEAD integrity}        & 12 & 16 \\
   \makecell{AES256-XTS-random, integrity HMAC-SHA256} & 16 & 32 \\
   \makecell{ChaCha20-random, integrity Poly1305}      & 16 & 32 \\
   \makecell{AEGIS128-random, AEAD integrity}          & 16 & 16 \\
   \makecell{AEGIS256-random, AEAD integrity}          & 16 & 32 \\
   \makecell{MORUS640-random, AEAD integrity}          & 16 & 16 \\
   \makecell{MORUS1280-random, AEAD integrity}         & 16 & 16 \\
   \bottomrule
  \end{tabular}
\caption{Tested encryption modes and metadata size.}
\label{tbl:modes}
\end{table}

\section{Performance}\label{sec:performance}
We present three simple performance benchmarks to investigate the usability
of our solution.
The first benchmark is a simple benchmark for linear access to the device
(read or write), the second is a synthetic I/O benchmark with interleaved reads
and writes, and the third is a C-code compilation time benchmark with
the underlying filesystem backed by our integrity protected device.

We used the mainline kernel 4.17~\cite{mainline_kernel} with additional
AEAD modules for AEGIS and MORUS~\cite{mosnacek_aead} and cryptsetup tool~\cite{luks}.
The tests ran on a basic installation of the Fedora Linux,
were repeated five times, and the arithmetic mean with the standard deviation
(in the format \emph{value $\pm$SD}) is then presented.
All cryptographic hardware acceleration modules were loaded.

The hardware configuration represents a typical desktop configuration that we
expect our solution would be used and is described in Table~\ref{tbl:hw_desktop}.

Tests on more older hardware (Lenovo laptop, see Table~\ref{tbl:hw_laptop})
was part of the abridged publication~\cite{fde_ifipsec18}. These tests used kernel 4.12
and do not include new AEAD modes implementations (not available at that time).

\begin{table}[h]
\centering
\def\arraystretch{1.3}
  \begin{tabular}{c}
   \toprule
   \makecell{\textbf{Hardware Configuration \#1}} \\
   \midrule
   \makecell{Intel Core i7-4790 3.60~GHz CPU (4 cores)} \\
    \makecell{CPU has AES-NI, SSE3 and AVX2 instructions} \\
    \makecell{32 GB RAM DDR3 1600~MHz} \\
   \midrule
    \makecell{Samsung SSD 850 Pro 512 GB, $2.5''$ Drive} \\
   \bottomrule
  \end{tabular}
\caption{Hardware configuration for desktop test.}
\label{tbl:hw_desktop}
\end{table}

\begin{table}[h]
\centering
\def\arraystretch{1.3}
  \begin{tabular}{c}
   \toprule
   \makecell{\textbf{Hardware Configuration \#2}} \\
   \midrule
    \makecell{Lenovo x240 laptop} \\
    \makecell{Intel Core i7-4600U 2.10~GHz CPU (4 cores)} \\
    \makecell{CPU has AES-NI and SSE3 instructions} \\
    \makecell{8 GB RAM Micron DDR3 1600~MHz} \\
   \midrule
    \makecell{Toshiba THNSFJ256GCSU 256 GB SSD, $2.5''$ Drive} \\
   \bottomrule
  \end{tabular}
\caption{Hardware configuration for older laptop test.}
\label{tbl:hw_laptop}
\end{table}

\subsection{Linear Access}
We ran tests that measured the speed of reads and writes to the entire device.
The I/O block size was fixed to 4~kB (presenting a page cache I/O
that always produces a page-sized I/O as the minimum).

\begin{figure}[H]
\includegraphics[width=\linewidth]{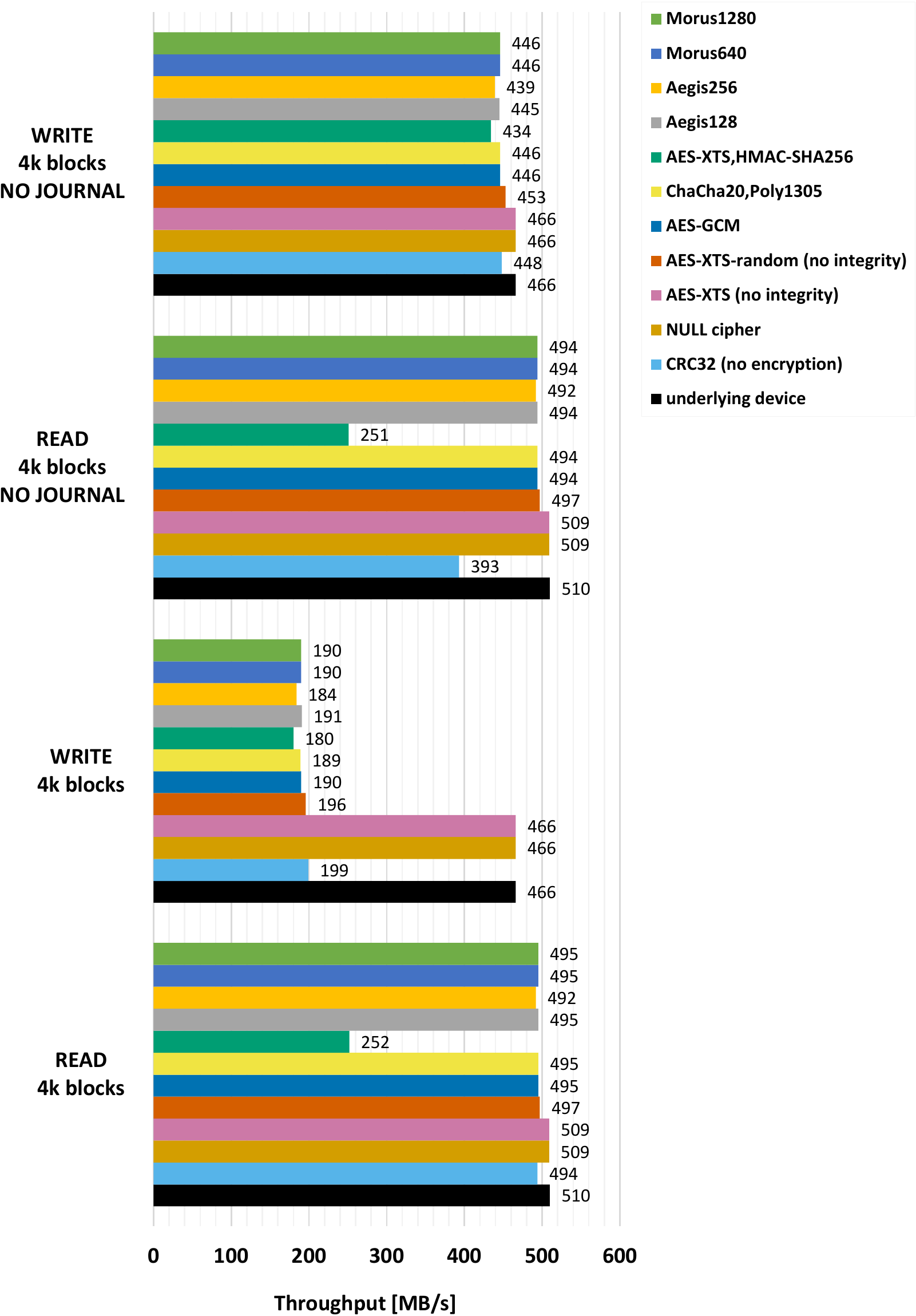}
\centering
\caption{Device throughput on a solid-state disk (desktop).}
\label{fig:speed-ssd}
\end{figure}

\begin{figure}[H]
\includegraphics[width=0.9\linewidth]{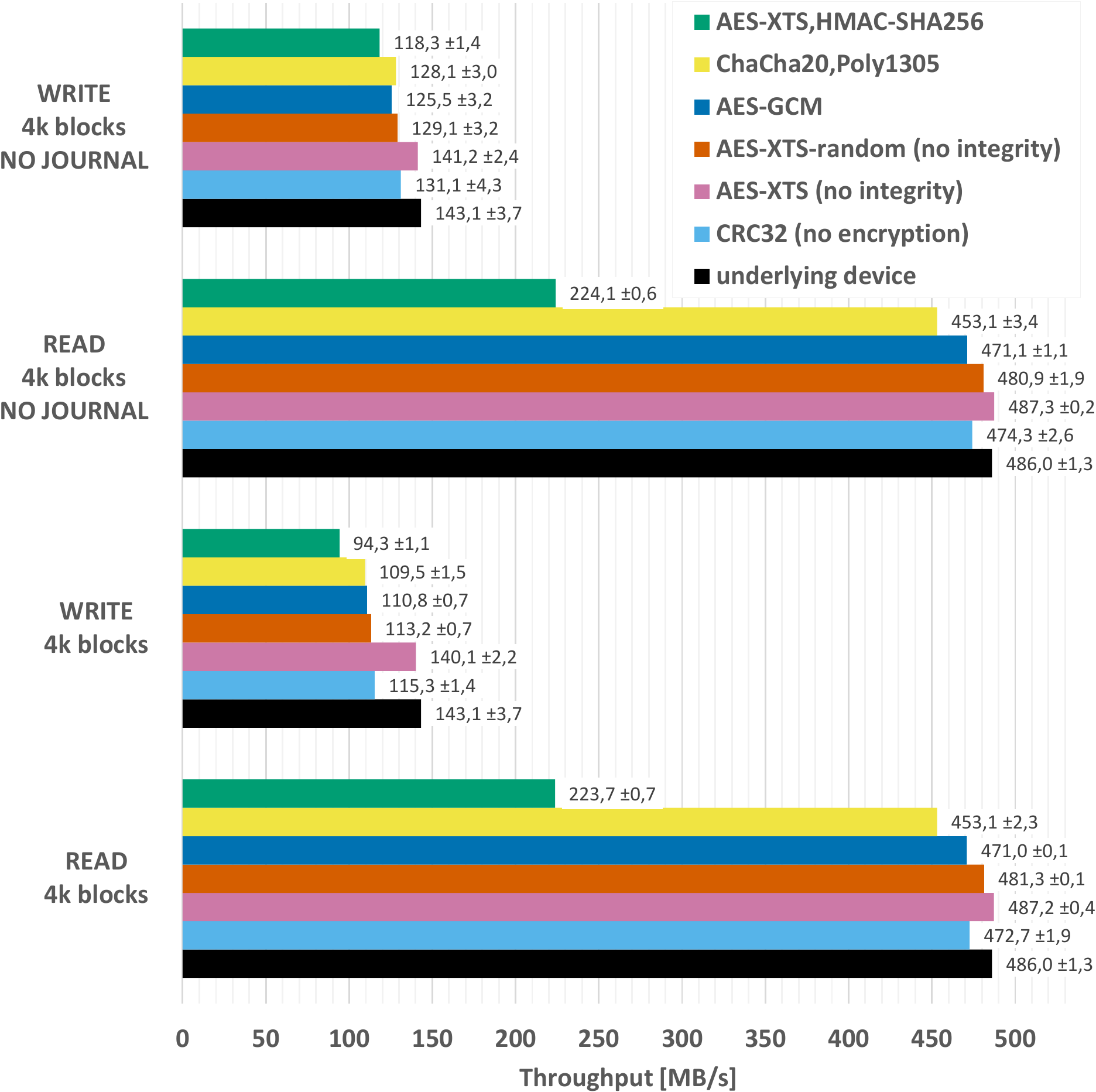}
\centering
\caption{Device throughput on a laptop solid-state disk.}
\label{fig:speed-x240}
\end{figure}

We ran the tests both with the \emph{dm-integrity}
data journal and without the journal (this simulates a situation when
a journaling is already present on an upper layer).
The measured data are presented in Figure~\ref{fig:speed-ssd} and Figure~\ref{fig:speed-x240}.

The measured data show that the overhead of the length-preserving encryption
in this scenario is almost negligible. The only visible overhead
is for the write operation.

Disabling the data journal has an effect only for write operations.
The most visible output is the overhead of additional metadata processing.
The overhead of cryptographic data integrity processing is visible
for all authenticated modes.

\subsection{Random I/O Throughput}
We simulated workload with the \emph{fio}~\cite{fio_github}
(Flexible I/O Tester) utility.
We used a \emph{fio} profile simulating mixed reads (70\%) and writes (30\%)
that generates random data access offsets with I/O operations of the 8k block
in size. The test uses 16 parallel jobs, asynchronous I/Os and runs for
at least 100 seconds.

The exact parameters of the \emph{fio} command were as follows:

\lstset{basicstyle=\ttfamily,showstringspaces=false}
\begin{lstlisting}[language=bash]
fio --filename=<DEVICE>
 --direct=0 --bs=8k --rw randrw:16
 --refill_buffers --norandommap
 --randrepeat=0 --ioengine=libaio
 --rwmixread=70 --iodepth=16
 --numjobs=16 --runtime=100
\end{lstlisting}

\begin{figure}[H]
\includegraphics[width=\linewidth]{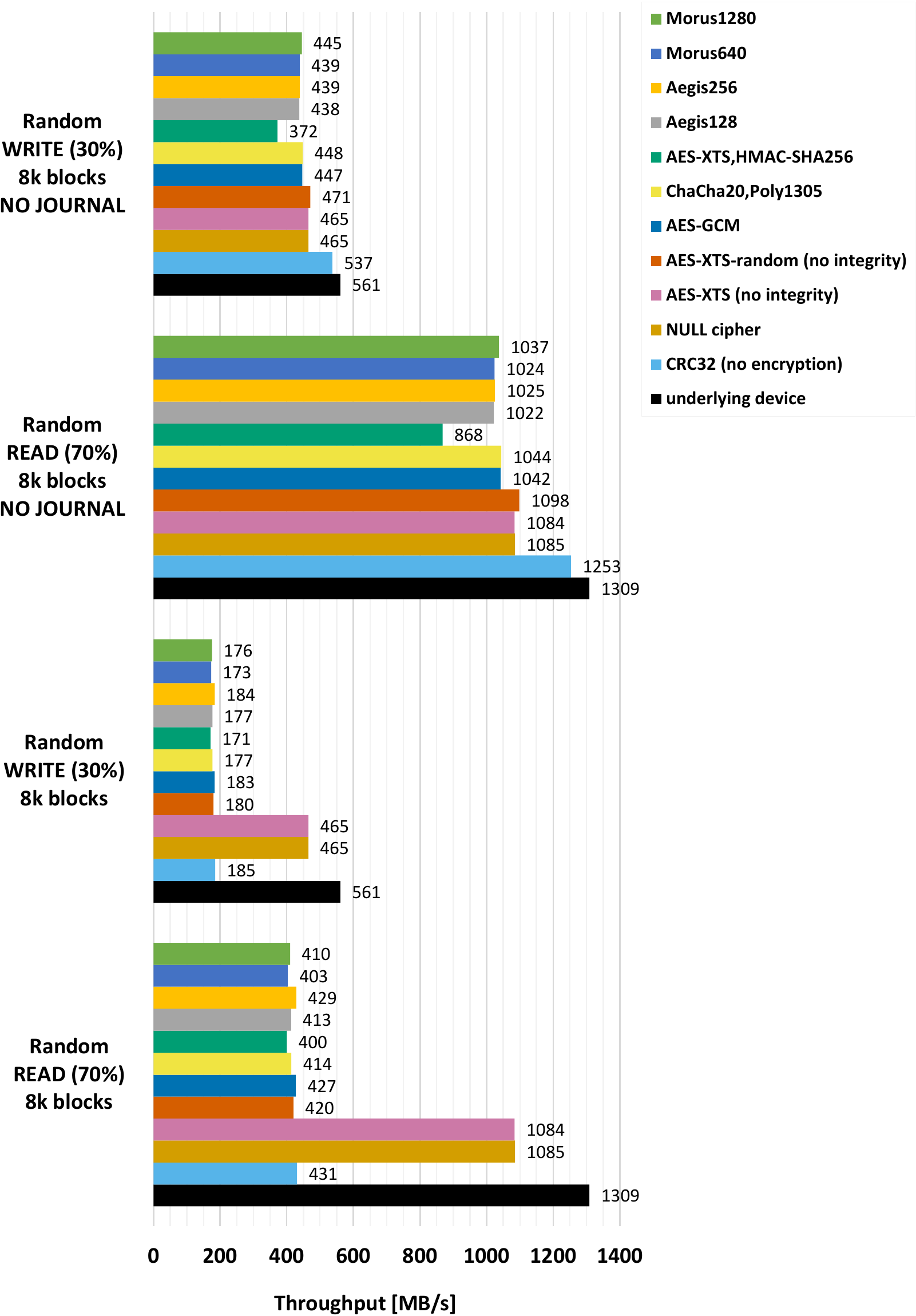}
\centering
\caption{\emph{fio} simulated load on a solid-state disk (desktop).}
\label{fig:fio-ssd}
\end{figure}

\begin{figure}[H]
\includegraphics[width=0.9\linewidth]{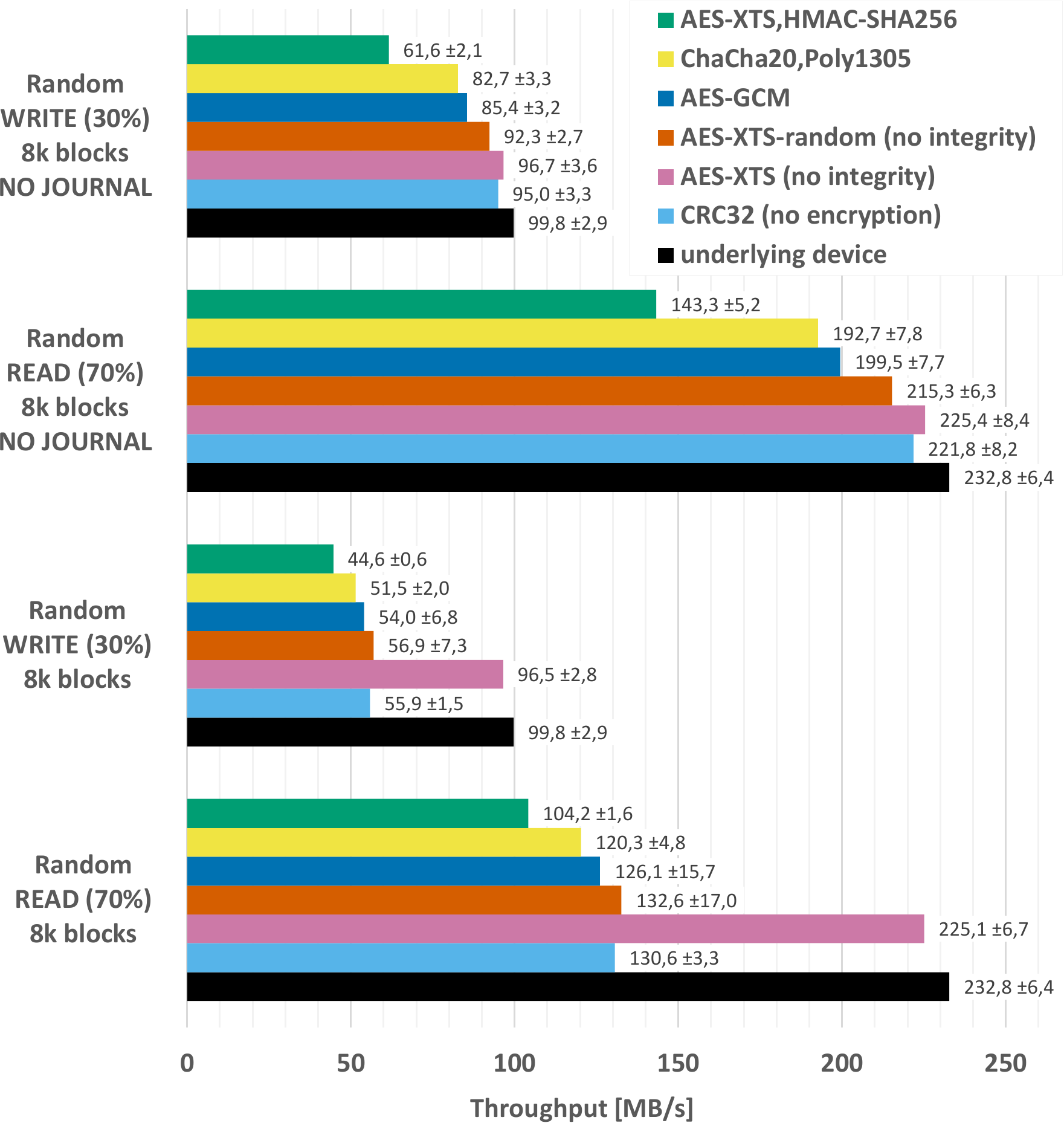}
\centering
\caption{\emph{fio} simulated load on a laptop solid-state disk.}
\label{fig:fio-x240}
\end{figure}

This test should represent the worst case scenario in our comparison.
The measured results are presented in Figures~\ref{fig:fio-ssd} and~\ref{fig:fio-x240}.
We can see that a journal has a small effect in this scenario; the major
visible slowdown is in the additional metadata processing.

\subsection{Compilation Time Test}
This test simulates a very simple application workload and measures
the overhead of an integrity-protected storage in a loaded system (common
laptop use).

The test unpacks mainline Linux kernel sources, flushes memory caches and
measures the compilation time of a specific kernel configuration (compilation
uses all available CPU cores in parallel).
We used the \emph{xfs} filesystem with default parameters over our device
in this scenario.

The measured times are presented in Table~\ref{tbl:comptimes}.
We also measured the same encryption mode with the \emph{dm-integrity}
journal switched off.

\begin{table}[h]
\def\arraystretch{2.2}
\centering
\begin{tabular}{@{} lcc @{}}
\toprule
\textbf{\makecell{Encryption mode\\and\\Authentication mode}} &
\textbf{\makecell{SSD\\\lbrack seconds\rbrack}} &
\textbf{\makecell{SSD\\(no journal)\\\lbrack seconds\rbrack}} \\
\midrule
\makecell{underlying device}                     & \makecell{1356,2 $\pm$40,9} & \\
\makecell{CRC32 checksum}                        & \makecell{1343,4 $\pm$13,9} & \makecell{1346,6 $\pm$31,7} \\
\makecell{NULL cipher}                           & \makecell{1365,5 $\pm$39,9} & \\
\makecell{AES-XTS-plain64}                       & \makecell{1341,6 $\pm$11,4} & \\
\makecell{AES-XTS-random}                        & \makecell{1352,9 $\pm$26,8} & \makecell{1335,1 $\pm$24,1} \\
\makecell{AES-GCM-random\\integrity AEAD}        & \makecell{1347,8 $\pm$26,4} & \makecell{1358,2 $\pm$55,8} \\
\makecell{AES-XTS-random\\integrity HMAC-SHA256} & \makecell{1343,8 $\pm$7,3}  & \makecell{1330,0 $\pm$19,2} \\
\makecell{ChaCha20-random\\integrity Poly1305}   & \makecell{1364,5 $\pm$38,0} & \makecell{1370,5 $\pm$39,2} \\
\makecell{AEGIS128-random\\integrity AEAD}       & \makecell{1353,5 $\pm$59,6} & \makecell{1339,7 $\pm$47,2} \\
\makecell{AEGIS256-random\\integrity AEAD}       & \makecell{1342,7 $\pm$35,4} & \makecell{1363,7 $\pm$36,3} \\
\makecell{MORUS640-random\\integrity AEAD}       & \makecell{1355,4 $\pm$16,9} & \makecell{1361,1 $\pm$15,8} \\
\makecell{MORUS1280-random\\integrity AEAD}      & \makecell{1344,5 $\pm$18,7} & \makecell{1340,6 $\pm$21,5} \\
\bottomrule
\end{tabular}
\caption{Compilation time test for desktop test.}
\label{tbl:comptimes}
\end{table}

From the measured numbers, we can see that the slowdown varies in all cases under 10\%.
Neither data journal nor authenticated encryption presents a significant slowdown
with this type of workload.

\section{Conclusions}\label{sec:conclusion}

Our goal was not only to show that the combination of confidentiality
and cryptographic data integrity protection is possible at the FDE layer,
but also to highlight the need for proper cryptographic data integrity
protection in general.
We focused on existing COTS devices without any specific hardware requirements.
Almost all laptops are currently delivered with an SSD. In this scenario,
the performance evaluation shows that our data integrity protection is usable
for these systems.

Our solution is based on existing concepts only, even though the proposed
authenticated encryption request format is designed to be compliant with
the IEEE 1619.1 standard~\cite{4523925}. This helps the deployment of our
integrity protection to systems that require security certifications.

The \emph{dm-integrity} module provides a generic solution for additional
metadata per sector by emulating HW-based DIF profiles with a software-defined
function.
The price to pay is decreased storage performance and storage capacity, but
we believe this is balanced by a zero-cost investment in required hardware
(our solution just uses existing COTS devices).

Our practical implementation has shown that the performance is in many
situations already acceptable for recent systems with SSDs
and could be in a limited scope also for devices with rotational drives.

The optimization of the \emph{dm-crypt} FDE implementation in Linux took
several years until it achieved today’s performance; thus, we expect that
the combination with the \emph{dm-integrity} will follow a similar path.

The extension to \emph{dm-crypt} FDE is also, as it was a major design goal,
algorithm-agnostic. The configuration of another encryption mode is just
a configuration option.

As a parallel effort~\cite{mosnacek_aead} to our work, an implementation
of new AEAGIS~\cite{aegis} and MORUS~\cite{morus} authenticated modes for Linux kernel
was finished, and we can say that the authentication encryption algorithm of choice today
is one of these CAESAR~\cite{caesar} finalists. Even if there is a better AEAD mode
introduced later (like a new GCM-SIV~\cite{cryptoeprint:2015:102,irtf-cfrg-gcmsiv}),
it can be easily used with our solution once it becomes available.
Both configurable metadata per-sector and encryption upgrades of algorithms
are the major advantages to hardware-based encryption solutions, where any
reconfiguration during their lifetime is almost impossible.

All code presented in this work is released under the open-source GPL2 license
and has been included in the Linux mainline kernel since the version 4.12
(new AEAD modes since the version 4.18) and userspace cryptsetup tool~\cite{luks}
since the version 2.0.0.

\subsection{Future Work}
Integrity protection is generally used on a higher layer than device
sectors~\cite{zhang2010end}.
The \emph{dm-integrity} feature could be used later even for another task such
as an application-level integrity protection (application could send its own
data to be stored on per-sector level) of storing additional Forward Error
Correction codes (the storage could then not only detect integrity problems
but also fix basic random data corruptions).

Additionally, the same principle can be applied to new storage device types,
such as persistent memory (or any byte-addressable persistent storage),
where we can easily increase the size of the additional authentication tag
(in principle, we can use virtual sector of any size).
In this case, the \emph{dm-integrity} layer can be omitted (the atomicity of
data and metadata writes is provided by the storage itself),
while the \emph{dm-crypt} cryptographic part of the solution remains the same.

\subsubsection*{Acknowledgments.}
The authors thank Arno Wagner, John Strunk, Ondrej Mosnáček, Virgil Gligor
and Ric Wheeler for valuable comments.

{
\footnotesize
\bibliographystyle{unsrt}
\bibliography{fde_aead}
}

\end{document}